\documentclass[conference]{IEEEtran}
\IEEEoverridecommandlockouts
\usepackage{cite}
\usepackage{amsmath,amssymb,amsfonts}
\usepackage{algorithmic}
\usepackage{graphicx}
\usepackage{textcomp}
\usepackage{xcolor}
\usepackage{threeparttable}
\usepackage{tabularx}
\usepackage{booktabs}
\usepackage{multirow}
\usepackage{caption}
\usepackage{colortbl}
\usepackage{subfigure}
\usepackage{pifont}
\usepackage[square,numbers,sort&compress]{natbib}

\usepackage[ruled,vlined]{algorithm2e}
\def\BibTeX{{\rm B\kern-.05em{\sc i\kern-.025em b}\kern-.08em
    T\kern-.1667em\lower.7ex\hbox{E}\kern-.125emX}}
\begin{document}

\title{
TCIM: \underline{T}riangle \underline{C}ounting Acceleration With Processing-\underline{I}n-\underline{M}RAM Architecture}

\author{Xueyan Wang\IEEEauthorrefmark{1}\IEEEauthorrefmark{3}, Jianlei Yang\IEEEauthorrefmark{2}\IEEEauthorrefmark{3}, Yinglin Zhao\IEEEauthorrefmark{1}, Yingjie Qi\IEEEauthorrefmark{2}, Meichen Liu\IEEEauthorrefmark{2}, Xingzhou Cheng\IEEEauthorrefmark{2},\\ Xiaotao Jia\IEEEauthorrefmark{1}\IEEEauthorrefmark{3}, Xiaoming Chen\IEEEauthorrefmark{4}, Gang Qu\IEEEauthorrefmark{5} and Weisheng Zhao\IEEEauthorrefmark{1}\IEEEauthorrefmark{3}\\
\IEEEauthorblockA{\IEEEauthorrefmark{1}Fert Beijing Research Institute, School of Microelectronics, Beihang University, Beijing, China
}
\IEEEauthorblockA{\IEEEauthorrefmark{2}School of Computer Science and Engineering, Beihang University, Beijing, China}
\IEEEauthorblockA{\IEEEauthorrefmark{3}Beijing Advanced Innovation Center for Big Data and Brain Computing, Beihang University, Beijing, China
}
\IEEEauthorblockA{\IEEEauthorrefmark{4}Institute of Computing Technology, Chinese Academy of Sciences, Beijing, China}
\IEEEauthorblockA{\IEEEauthorrefmark{5}Department of Electrical and Computer Engineering, University of Maryland, College Park, MD, USA}
Email: jianlei@buaa.edu.cn~~weisheng.zhao@buaa.edu.cn
\thanks{X. Wang, J. Yang, X. Jia and W. Zhao's work are supported in part by the National Natural Science Foundation of China (61602022), State Key Laboratory of Computer Architecture (CARCH201917), NSFC 61701013, State Key Laboratory of Software Development Environment (SKLSDE-2018ZX-07), National Key Technology Program of China (2017ZX01032101), CCF-Tencent IAGR20180101 and the 111 Talent Program B16001. X. Chen's work was supported by Beijing Academy of Artificial Intelligence (BAAI).}
}

\maketitle

\begin{abstract}

Triangle counting (TC) is a fundamental problem in graph analysis and has found numerous applications, which motivates many TC acceleration solutions in the traditional computing platforms like GPU and FPGA. However, these approaches suffer from the bandwidth bottleneck because TC calculation involves a large amount of data transfers.
In this paper, we propose to overcome this challenge by designing a TC accelerator utilizing the emerging processing-in-MRAM (PIM) architecture. The true innovation behind our approach is a novel method to perform TC with bitwise logic operations (such as \texttt{AND}), instead of the traditional approaches such as matrix computations. This enables the efficient in-memory implementations of TC computation, which we demonstrate in this paper with computational Spin-Transfer Torque Magnetic RAM (STT-MRAM) arrays. Furthermore, we develop
customized graph slicing and mapping techniques to speed up the computation and reduce the energy consumption.
We use a device-to-architecture co-simulation framework to validate our proposed TC accelerator. The results show that our data mapping strategy could reduce $99.99\%$ of the computation and $72\%$ of the memory \texttt{WRITE} operations.
Compared with the existing GPU or FPGA accelerators, our in-memory accelerator achieves speedups of $9\times$ and $23.4\times$, respectively, and a $20.6\times$ energy efficiency improvement over the FPGA accelerator.

\end{abstract}


\section{Introduction}

Triangles are the basic substructure of networks and play critical roles in network analysis.
Due to the importance of triangles, triangle counting problem (TC), which counts the number of triangles in a given graph, is essential for analyzing networks and generally considered as the first fundamental step in calculating metrics such as clustering coefficient and transitivity ratio, as well as other tasks such as community discovery, link prediction, and Spam filtering \cite{tcReview}.
TC problem is not hard but they are all memory bandwidth intensive thus time-consuming. As a result, researchers from both academia and industry have proposed many TC acceleration methods ranging from sequential to parallel, single-machine to distributed, and exact to approximate.
From the computing hardware perspective, these acceleration strategies are generally executed on CPU, GPU or FPGA, and are based on Von-Neumann architecture \cite{tcReview,XiongTCCPGPU,XiongTCFPGA}.
However, due to the fact that most graph processing algorithms have low computation-memory ratio and high random data access patterns, there are frequent data transfers between the computational unit and memory components which consumes a large amount of time and energy.

In-memory computing paradigm performs computation where the data resides. It can save most of the off-chip data communication energy and latency by exploiting the large internal memory inherent bandwidth and inherent parallelism \cite{MutluDRAM,DBLP:conf/dac/LiXZZLX16}. As a result, in-memory computing has appeared as a viable way to carry out the computationally-expensive and memory-intensive tasks \cite{LiBingOverview,FanAligns}.
This becomes even more promising when being integrated with the emerging non-volatile STT-MRAM memory technologies. This integration, called Processing-In-MRAM (PIM), offers fast write speed, low write energy, and high write endurance among many other benefits \cite{wang2018current,DBLP:journals/tvlsi/JainRRR18}.

In the literature, there have been some explorations on in-memory graph algorithm accelerations \cite{ChenHPCA,FanGraphs,WangYuASPDAC,QianMicro}, however, existing TC algorithms, including the intersection-based and the matrix multiplication-based ones, cannot be directly implemented in memory. For large sparse graphs, highly efficient PIM architecture, efficient graph data compression and data mapping mechanisms are all critical for the efficiency of PIM accelerations. Although there are some compression methods for sparse graph, such as compressed sparse column (CSC), compressed sparse row (CSR), and coordinate list (COO) \cite{ChenHPCA}, these representations cannot be directly applied to in-memory computation either.
In this paper, we propose and design the first in-memory TC accelerator that overcomes the above barriers.
Our main contributions can be summarized as follows:

\begin{itemize}

\item We propose a novel TC method that uses massive bitwise operations to enable in-memory implementations.
\item We propose strategies for data reuse and exchange, and data slicing for efficient graph data compression and mapping onto in-memory computation architectures.
\item We build a TC accelerator with the sparsity-aware processing-in-MRAM architecture. A device-to-architecture co-simulation demonstrates highly encouraging results.

\end{itemize}

The rest of the paper is organized as follows:
Section~\ref{sec:preliminary} provides some preliminary knowledge of TC and in-memory computing.
Section~\ref{sec:tc} introduces the proposed TC method with bitwise operations, and Section~\ref{sec:pimArch} elaborates a sparsity-aware processing-in-MRAM architecture which enables highly efficient PIM accelerations.
Section~\ref{sec:exper} demonstrates the experimental results and Section~\ref{sec:conclusion} concludes.

\section{Preliminary}\label{sec:preliminary}

\subsection{Triangle Counting}

Given a graph, triangle counting (TC) problem seeks to determine the number of triangles. The sequential algorithms for TC can be classified into two groups.
In the {matrix multiplication based algorithms}, a triangle is a closed path of length three, namely a path of three vertices begins and ends at the same vertex. If $A$ is the adjacency matrix of graph $G$, $A^3[i][i]$ represents the number of paths of length three beginning and ending with vertex $i$. Given that a triangle has three vertices and will be counted for each vertex, and the graph is undirected (that is, a triangle $i-p-q-i$ will be also counted as $i-q-p-i$), the number of triangles in $G$ can be obtained as $trace(A^3)/6$, where $trace$ is the sum of elements on the main diagonal of a matrix.
In the {set intersection based algorithms}, it iterates over each edge and finds common elements from adjacency lists of head and tail nodes.
A lot of CPU, GPU and FPGA based optimization techniques have been proposed \cite{tcReview,XiongTCCPGPU,XiongTCFPGA}. These works show promising results of accelerating TC, however, these strategies all suffer from the performance and energy bottlenecks brought by the significant amount of data transfers in TC.

\subsection{In-Memory Computing with STT-MRAM}
STT-MRAM is a promising candidate for the next generation main memory because of its properties such as near-zero leakage, non-volatility, high endurance, and compatibility with the CMOS manufacturing process \cite{wang2018current}. In particular, prototype STT-MRAM chip demonstrations and commercial MRAM products have been available by companies such as Everspin and TSMC.
STT-MRAM stores data with magnetic-resistances instead of conventional charge based store and access. This enables MRAM to provide inherent computing capabilities for bitwise logic with minute changes to peripheral circuitry \cite{DBLP:journals/tvlsi/JainRRR18}\cite{yang2018exploiting}.

\begin{figure}[t]
\centering
\includegraphics[width = 0.9\linewidth]{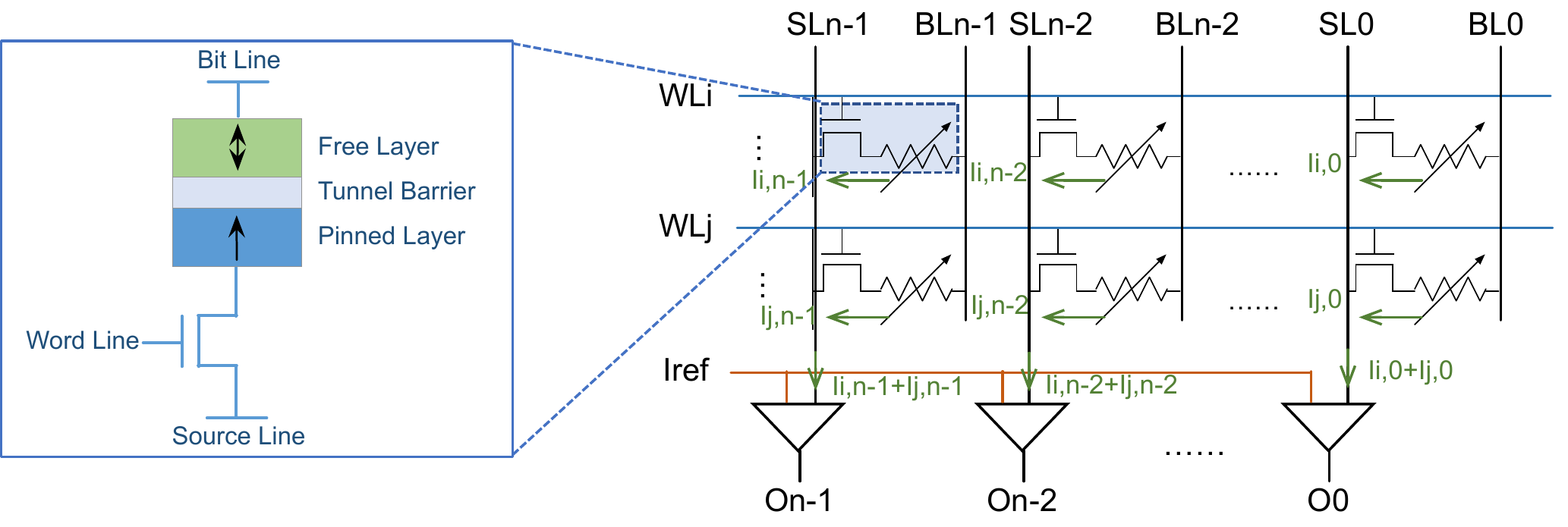}
\caption{Typical STT-MRAM bit-cell and paradigm of computing in STT-MRAM array.}
\label{fig:cim}
\end{figure}

As the left part of Fig.~\ref{fig:cim} shows, a typical STT-MRAM bit-cell consists of an access transistor and a Magnetic Tunnel Junction (MTJ), which is controlled by bit-line (BL), word-line (WL) and source-line (SL).
The relative magnetic orientations of pinned ferromagnetic layer (PL) and free ferromagnetic layer (FL) can be stable in parallel (\texttt{P} state) or anti-parallel (\texttt{AP} state), corresponding to low resistance ($R_{\rm P}$) and high resistance ($R_{\rm AP}$, $R_{\rm AP}>R_{\rm P}$), respectively.
\texttt{READ} operation is done by enabling WL signal, applying a voltage $V_{\rm read}$ across BL and SL, and sensing the current that flows ($I_{\rm P}$ or $I_{AP}$) through the MTJ. By comparing the sense current with a reference current ($I_{\rm ref}$,), the data stored in MTJ cell (logic `0' or logic `1') could be readout.
\texttt{WRITE} operation can be performed by enabling WL, then applying an appropriate voltage ($V_{\rm write}$) across BL and SL to pass a current that is greater than the critical MTJ switching current.
To perform bitwise logic operation, as demonstrated in the right part of Fig.~\ref{fig:cim}, by simultaneously enabling $WL_i$ and $WL_j$, then applying $V_{\rm read}$ across $BL_k$ and $SL_k$ ($k \in [0,n-1]$), the current that feeds into the $k$-th sense amplifier (SA) is a summation of the currents flowing through $MTJ_{i,k}$ and $MTJ_{j,k}$, namely $I_{i,k}+I_{j,k}$.
With different reference sensing current, various logic functions of the enabled word line can be implemented.

\section{Triangle Counting with Bitwise Operations}\label{sec:tc}

In this section, we seek to perform TC with massive bitwise operations, which is the enabling technology for in-memory TC accelerator.
Let $A$ be the adjacency matrix representation of a undirected graph $G(V,E)$, where  $A[i][j]\in \{0,1\}$ indicates whether there is an edge between vertices $i$ and $j$.
If we compute $A^2=A*A$, then the value of $A^2[i][j]$ represents the number of distinct paths of length two between vertices $i$ and $j$.
In the case that there is an edge between vertex $i$ and vertex $j$, and $i$ can also reach $j$ through a path of length two, where the intermediate vertex is $k$, then vertices $i$, $j$, and $k$ form a triangle.
As a result, the number of triangles in $G$ is equal to the number of non-zero elements ($nnz$) in $A \cap A^2$ (the symbol `$\cap$' defines element-wise multiplication here), namely
\begin{equation}\label{equ:eq1}
  TC(G)=nnz(A \cap A^2)
\end{equation}
Since $A[i][j]$ is either zero or one, we have
\begin{equation}\label{equ:eq2}
  (A\cap A^2)[i][j]=
  \begin{cases}
    0, & \text{if}\ A[i][j]=0;\\
    A^2[i][j], & \text{if}\ A[i][j]=1.
  \end{cases}
\end{equation}
According to Equation~(\ref{equ:eq2}),
\begin{equation}\label{equ:eq3}
\begin{split}
  nnz(A \cap A^2)&=\sum\sum\nolimits_{A[i][j]=1}A^2[i][j]\\
\end{split}
\end{equation}
Because the element in $A$ is either zero or one, the bitwise Boolean \texttt{AND} result is equal to that of the mathematical multiplication, thus

\begin{equation}\label{equ:eq4}
\begin{split}
  A^2[i][j]& =\sum_{k=0}^{n} A[i][k]*A[k][j]=\sum_{k=0}^{n} {AND}(A[i][k],A[k][j])\\
 & ={BitCount}({AND}(A[i][*],A[*][j]^T))
\end{split}
\end{equation}
in which \texttt{BitCount} returns the number of `1's in a vector consisting of `0' and `1', for example, $BitCount(0110)=2$.

Combining equations ~(\ref{equ:eq1}), (\ref{equ:eq3}) and (\ref{equ:eq4}), we have
\begin{equation}
\begin{split}
  TC(G)&={BitCount}({AND}(A[i][*],A[*][j]^T)),\\
  &\quad \text{in which }A[i][j]=1
  \end{split}
\end{equation}

Therefore, TC can be completed by only \texttt{AND} and \texttt{BitCount} operations (massive for large graphs).
Specifically, for each non-zero element $A[i][j]=1$, the $i$-th row ($R_i=A[i][*]$) and the $j$-th column ($C_j=A[*][j]^T$) are executed \texttt{AND} operation, then the \texttt{AND} result is sent to a bit counter module for accumulation.
Once all the non-zero elements are processed as above, the value in the accumulated \texttt{BitCount} is exactly the number of triangles in the graph.

\begin{figure}[htb]
\centering
\includegraphics[width = 0.85\linewidth]{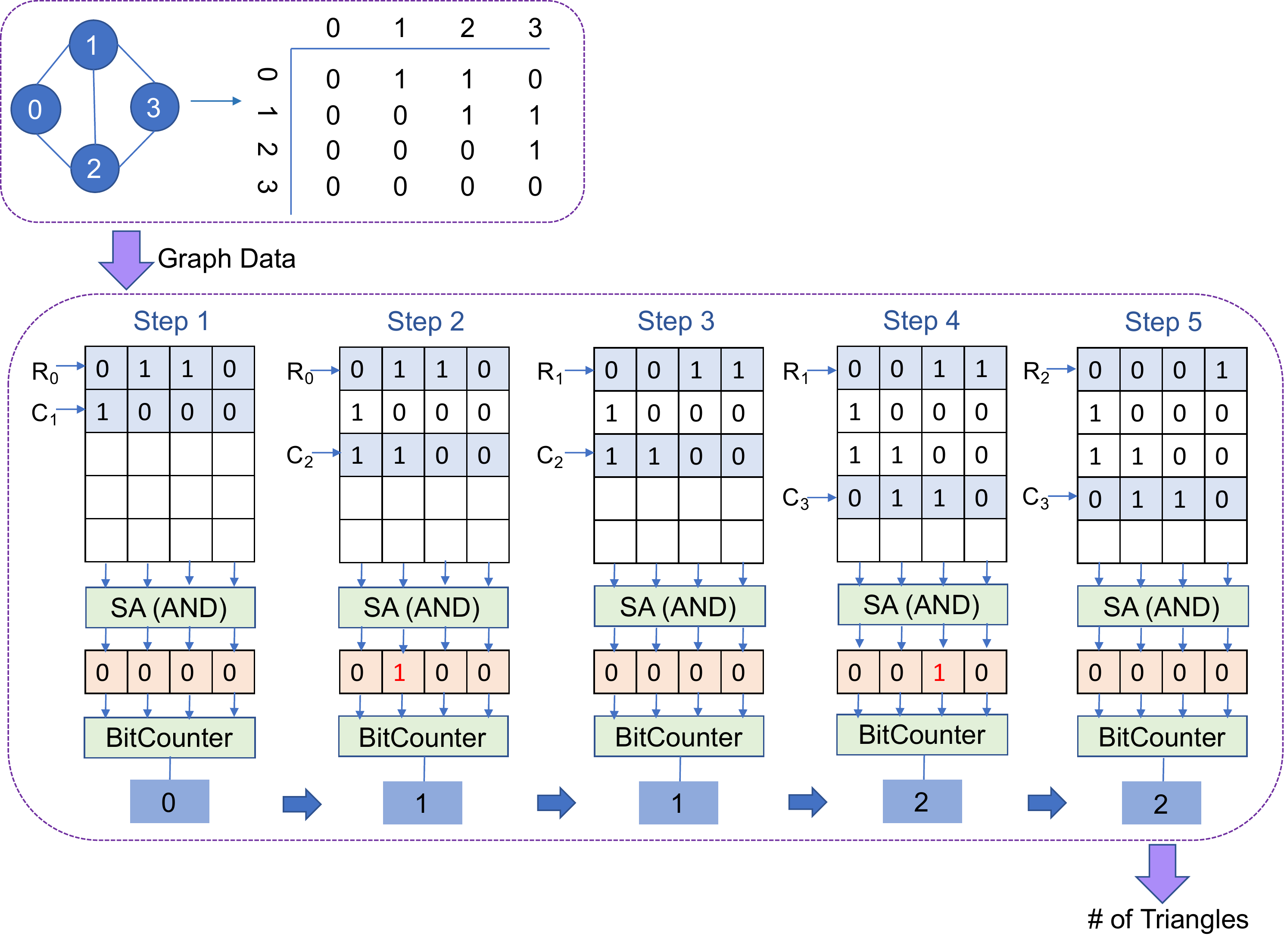}
\caption{Demonstrations of triangle counting with \texttt{AND} and \texttt{BitCount} bit-wise operations.}
\label{fig:TCProcedure}
\end{figure}

Fig.~\ref{fig:TCProcedure} demonstrates an illustrative example for the proposed TC method.
As the left part of the figure shows, the graph has four vertices, five edges and two triangles ($0-1-2-0$ and $1-2-3-1$), and the adjacency matrix is given.
The non-zero elements in $A$ are $A[0][1]$, $A[0][2]$, $A[1][2]$, $A[1][3]$, and $A[2][3]$.
For $A[0][1]$, row $R_0$=`0110' and column $C_1$=`1000' are executed with \texttt{AND} operation, then the \texttt{AND} result `0000' is sent to the bit counter and gets a result of zero. Similar operations are performed to other four non-zero elements.
After the execution of the last non-zero element $A[2][3]$ is finished, the accumulated \texttt{BitCount} result is two, thus the graph has two triangles.

The proposed TC method has the following advantages. First, it avoids the time-consuming multiplication. When the operation data are either zero or one, we can implement the multiplication with \texttt{AND} logic.
Second, the proposed method does not need to store the intermediate results that are larger than one (such as the elements in $A^2$), which are cumbersome to store and calculate.
Third, it does not need complex control logic.
Given the above three advantages, the proposed TC method is suitable for in-memory implementations.

\section{Sparsity-Aware Processing-In-MRAM Architecture}\label{sec:pimArch}

To alleviate the memory bottleneck caused by frequent data transfers in traditional TC algorithms, we implement an in-memory TC accelerator based on the novel TC method presented in the previous section.
Next, we will discuss several dataflow mapping techniques to minimize space requirements, data transfers and computation in order to accelerate the in-memory TC computation.

\subsection{Data Reuse and Exchange}

Recall that the proposed TC method iterates over each non-zero element in the adjacency matrix, and loads corresponding rows and columns into computational memory for \texttt{AND} operation, followed by a \texttt{BitCount} process. When the size of the computational memory array is given, it is important to reduce the unnecessary space and memory operations.
We observe that for \texttt{AND} computation, the non-zero elements in a row reuse the same row, and the non-zero elements in a column reuse the same column. The proposed data reuse mechanism is based on this observation.

Assume that the non-zero elements are iterated by rows, then the current processed row only needs to be loaded once, at the same time the corresponding columns are loaded in sequence.
Once all the non-zero elements in a row have been processed, this row will no longer be used in future computation, thus we can overwrite this row by the next row to be processed.
However, the columns might be used again by the non-zero elements from the other rows.
Therefore, before loading a certain column into memory for computation, we will first check whether this column has been loaded, if not, the column will be loaded to a spare memory space. In case that the memory is full, we need to select one column to be replaced with the current column. We choose the least recently used (LRU) column for replacement, and more optimized replacement strategy could be possible.

As demonstrated in Fig.~\ref{fig:TCProcedure}, in step $1$ and step $2$, the two non-zero elements $A[0][1]$ and $A[0][2]$ of row $R_0$ are processed respectively, and corresponding columns $C_1$ and $C_2$ are loaded to memory.
Next, while processing $A[1][2]$ and $A[1][3]$, $R_1$ will overlap $R_0$ and reuse existing $C_2$ in step $3$, and load $C_3$ in step $4$.
In step $5$, to process $A[2][3]$, $R_1$ will be overlapped by $R_2$, and $C_3$ is reused.
Overlapping the rows and reusing the columns can effectively reduce unnecessary space utilization and memory \texttt{WRITE} operations.

\subsection{Data Slicing}

To utilize the sparsity of the graph to reduce the memory requirement and unnecessary computation, we propose a data slicing strategy for graph data compression.

Assume $R_i$ is the $i$-th row, and $C_j$ is the $j$-th column of the adjacency matrix $A$ of graph $G(V,E)$. The slice size is $|S|$ (each slice contains $|S|$ bits), then each row and column has $\lceil\frac{|V|}{|S|}\rceil$ number of slices.
The $k$-th slice in $R_i$, which is represented as $R_i S_k$, is the set of $\{A[i][k*|S|],\cdots,A[i][(k+1)*|S|-1]$.
We define that slice $R_i S_k$ is \textbf{\textit{valid}} if and only if $\exists A[i][t] \in R_i S_k,A[i][t]=1,t\in [k*|S|,(k+1)*|S|-1]$.

Recall that in our proposed TC method, for each non-zero element in the adjacency matrix, we compute the \texttt{AND} result of the corresponding row and column.
With row and column slicing, we will perform the \texttt{AND} operation in the unit of slices. For each $A[i][j]=1$, we only process the valid slice pairs, namely only when both the row slice $R_i S_k$ and column slice $C_j S_k$ are valid, we will load the valid slice pair $(R_iS_k,C_jS_k)$ to the computational memory array and perform \texttt{AND} operation.

\begin{figure}[htbp]
\centering
\includegraphics[width = 0.82\linewidth]{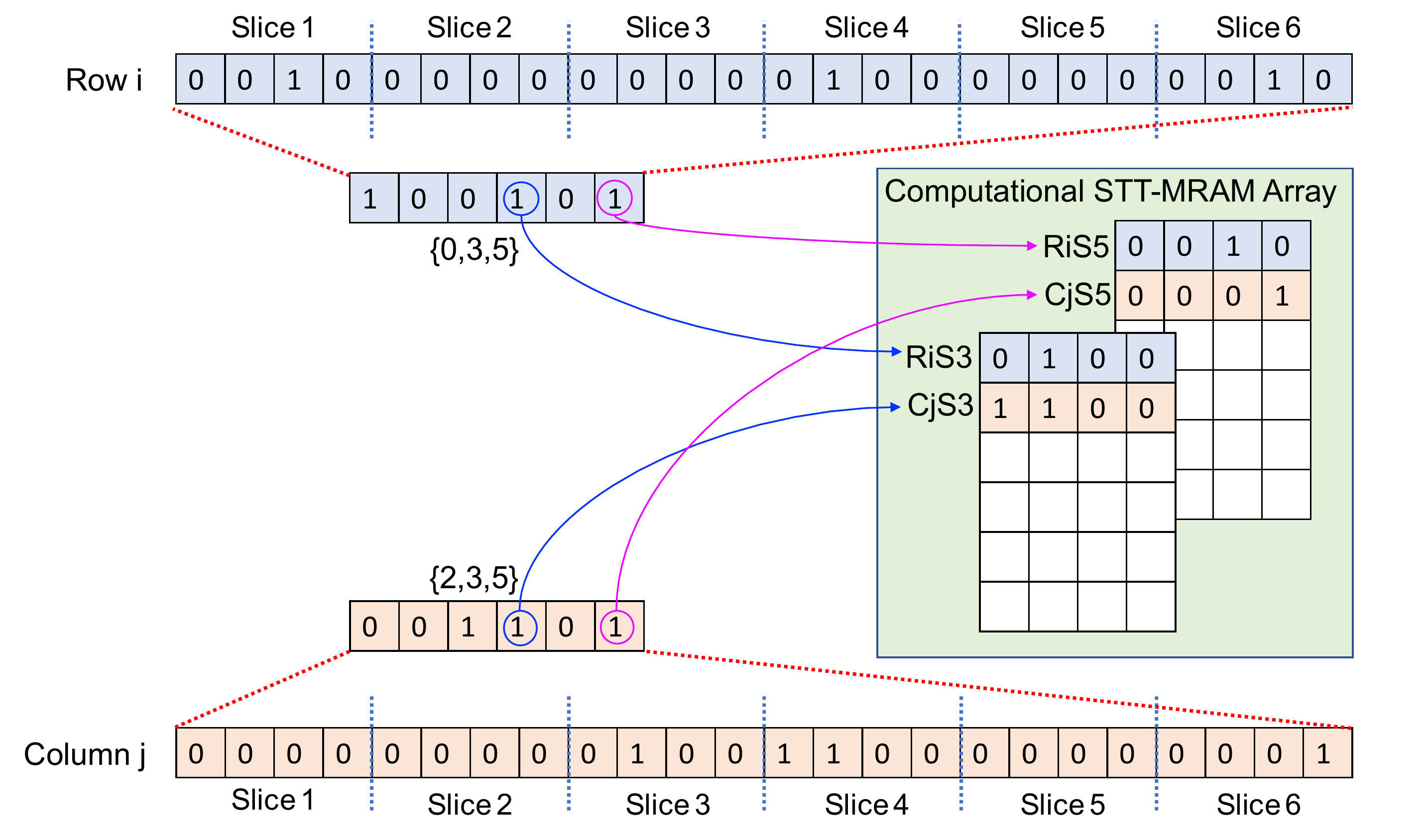}
\caption{Sparsity-aware data slicing and mapping.}
\label{fig:rowslicing}
\end{figure}

Fig.~\ref{fig:rowslicing} demonstrates an example, after row and column slicing, only slice pairs $(R_iS_3,C_jS_3)$ and $(R_iS_5,C_jS_5)$ are valid, therefore, we only load these slices for \texttt{AND} computation. This scheme can reduce the needed computation significantly, especially in the large sparse graphs.

\textit{Memory requirement of the compressed graph data.}
With the proposed row and column slicing strategy, we need to store the index of valid slices and the detailed data information of these slices.
Assuming that the number of valid slices is $N_{VS}$, the slice size is $|S|$, and we use an integer (four Bytes) to store each valid slice index, then the needed space for overall valid slice index is $IndexLength = N_{VS} \times 4$ Bytes.
The needed space to store the data information of valid slices is $DataLength = N_{VS} \times |S| / 8$ Bytes.
Therefore, the overall needed space for graph $G$ is $N_{VS} \times (|S| / 8 + 4)$ Bytes, which is determined by the sparsity of $G$ and the slice size.
In this paper, we set $|S|=64$ in the experimental result section.
Given that most graphs are highly sparse, the needed space to store the graph can be trivial. \textbf{Moreover, the proposed format of compressed graph data is friendly for directly mapping onto the computational memory arrays to perform in-memory logic computation.}

\begin{algorithm}[t]
\footnotesize
\KwIn{Graph $G(V,E)$.}
\KwOut{The number of triangles in $G$.}
$TC\_G$ = 0\;
Represent $G$ with adjacent matrix $A$\;
\For {each edge $e\in E$ with $A[i][j]=1$}{
Partition $R_i$ into slices\;
Partition $C_j$ into slices\;
\For {each valid slice pair ($R_iS_k$,$C_jS_k$)}{
$TC\_G$ += \textbf{COMPUTE} ($R_iS_k$,$C_jS_k$)\;
}
}
\textbf{return} $TC\_G$ as the number of triangles in $G$.\\
----------------------------------------\\
\textbf{COMPUTE} ($Slice1$, $Slice2$)
{\\
load $Slice1$ into memory\;
\If {$Slice2$ has not been loaded}{
\eIf {there is no enough space}{
Replace least recently used slice with $Slice2$\;
}{Load $Slice2$ into memory\;}
}
\textbf{return} $BitCount(AND(Slice1,Slice2))$.
}
\caption{TCIM: Triangle Counting with Processing-In-MRAM Architecture.}
\label{alg:dataMapping}
\end{algorithm}

\subsection{Processing-In-MRAM Architecture}

\begin{figure*}[t]
\centering
\includegraphics[width = 0.75\linewidth]{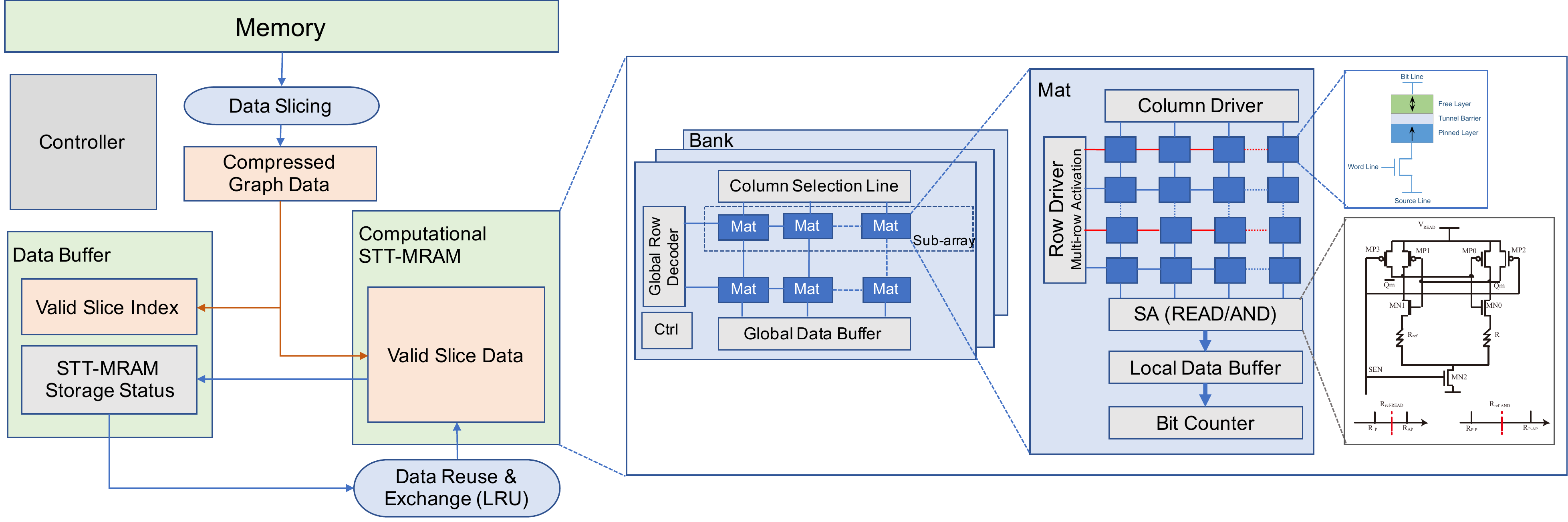}
\caption{Overall processing-in-MRAM architecture.}
\label{fig:overallArch}
\end{figure*}

Fig.~\ref{fig:overallArch} demonstrates the overall architecture of processing-in-MRAM.
The graph data will be sliced and compressed, and represented by the valid slice index and corresponding slice data.
According to the valid slice indexes in the data buffer, we load the corresponding valid slice pairs into computational STT-MRAM array for bitwise computation.
The storage status of STT-MRAM array (such as which slices have been loaded) is also recorded in the data buffer and utilized for data reuse and exchange.

As for the computational memory array organization, each chip consists of multiple Banks and works as computational array.
Each Bank is comprised of multiple computational memory sub-arrays, which are connected to a global row decoder and a shared global row buffer. Read circuit and write driver of the memory array are modified for processing bitwise logic functions. Specifically, the operation data are all stored in different rows in memory arrays. The rows associated with operation data will be activated simultaneously for computing. Sense amplifiers are enhanced with \texttt{AND} reference circuits to realize either \texttt{READ} or \texttt{AND} operations. By generating $R_\text{ref-AND}\in (R_\text{P-P},R_\text{P-AP})$, the output by the sense amplifier is the \texttt{AND} result of the data that is stored in the enabled WLs.

\subsection{Pseudo-code for In-Memory TC Acceleration}

Algorithm~\ref{alg:dataMapping} demonstrates the pseudo-code for TC accelerations with the proposed processing-in-MRAM architecture.
It iterates over each edge of the graph, partitions the corresponding rows and columns into slides, then loads the valid slice pairs onto computational memory for \texttt{AND} and \texttt{BitCount} computation. In case that there is no enough memory space, it adopts an LRU strategy to replace a least recently used slice.

\section{Experimental Results}\label{sec:exper}

\subsection{Experimental Setup}

To validate the effectiveness of the proposed approaches, comprehensive device-to-architecture evaluations along with two in-house simulators are developed.
At the device level, we jointly use the Brinkman model and Landau-Lifshitz-Gilbert (LLG) equation to characterize MTJ \cite{yang2015radiation}. The key parameters for MTJ simulation are demonstrated in Table~\ref{tab:parameter}.
For the circuit-level simulation, we design a Verilog-A model for 1T1R STT-MRAM device, and characterize the circuit with $45$nm FreePDK CMOS library.
We design a bit counter module based on Verilog HDL to obtain the number of non-zero elements in a vector. Specifically, we split the vector and feed each $8$-bit sub-vector into an $8$-$256$ look-up-table to get its non-zero element number, then sum up the non-zero numbers in all sub-vectors. We synthesis the module with Synopsis Tool and conduct post-synthesis simulation based on $45$nm FreePDK.
After getting the device level simulation results, we integrate the parameters in the open-source NVSim simulator \cite{NVSim} and obtain the memory array performance.
In addition, we develop a simulator in Java for the processing-in-MRAM architecture, which simulates the proposed function mapping, data slicing and data mapping strategies.
Finally, a behavioural-level simulator is developed in Java, taking architectural-level results and memory array performance to calculate the latency and energy that spends on TC in-memory accelerator.
To provide a solid comparison with other accelerators, we select from the real-world graphs from SNAP dataset \cite{snapnets} (see TABLE~\ref{tab:graphpara}), and run comparative baseline intersect-based algorithm on Inspur blade system with the Spark GraphX framework on Intel E5430 single-core CPU. Our TC in-memory acceleration algorithm also runs on single-core CPU, and the STT-MRAM computational array is set to be $16$ MB.

\begin{table}[htbp]
\setlength{\tabcolsep}{14pt}
\footnotesize
\caption{Key parameters for MTJ simulation.}
\label{tab:parameter}
\centering
\begin{tabular}{l|l}
\specialrule{0.8pt}{0pt}{0pt}
 Parameter & Value \\ \hline
 MTJ Surface Length & $40$ $nm$ \\
 MTJ Surface Width & $40$ $nm$ \\
 Spin Hall Angle & $0.3$ \\
 Resistance-Area Product of MTJ & $10^{-12}$ $\Omega \cdot m^2$ \\
 Oxide Barrier Thickness & $0.82$ $nm$ \\
 TMR & $100\%$ \\
 Saturation Field & $10^6$ $A/m$ \\
 Gilbert Damping Constant & $0.03$ \\
 Perpendicular Magnetic Anisotropy & $4.5 \times 10^5$ $A/m$ \\
 Temperature & $300 K$ \\
\specialrule{0.8pt}{0pt}{0pt}
\end{tabular}
\end{table}

\begin{table}[t]
\setlength{\tabcolsep}{10pt}
\footnotesize
\caption{Selected graph dataset.}
\label{tab:graphpara}
\centering
\begin{tabular}{l|rrr}
\specialrule{0.8pt}{0pt}{0pt}
 Dataset & \# Vertices & \# Edges & \# Triangles \\ \hline
 ego-facebook & 4039 & 88234 & 1612010 \\
 email-enron & 36692 & 183831 & 727044 \\
 com-Amazon & 334863 & 925872 & 667129 \\
 com-DBLP & 317080 & 1049866 & 2224385 \\
 com-Youtube & 1134890 & 2987624 & 3056386 \\
 roadNet-PA & 1088092 & 1541898 & 67150 \\
 roadNet-TX & 1379917 & 1921660 & 82869 \\
 roadNet-CA & 1965206 & 2766607 & 120676 \\
 com-LiveJournal & 3997962 & 34681189 & 177820130 \\
\specialrule{0.8pt}{0pt}{0pt}
\end{tabular}
\end{table}

\subsection{Benefits of Data Reuse and Exchange}

TABLE~\ref{tab:sliceDataSize} shows the memory space required for the bitwise computation. For example, the largest graph {\it com-lj} will need $16.8$ MB without incurring any data exchange. On average, only $18$ KB per $1000$ vertices is needed for in-memory computation.

\begin{table}[t]
\setlength{\tabcolsep}{6pt}
\footnotesize
\caption{Valid slice data size (MB).}
\label{tab:sliceDataSize}
\centering
\begin{tabular}{lr|lr|lr}
\specialrule{0.8pt}{0pt}{0pt}
ego-facebook & 0.182 & com-DBLP & 7.6 & roadNet-TX & 12.38 \\
email-enron & 1.02 & com-Youtube & \bf{16.8} & roadNet-CA & \bf{16.78} \\
com-Amazon & 7.4 & roadNet-PA & 9.96 & com-lj & \bf{16.8} \\
\specialrule{0.8pt}{0pt}{0pt}
\end{tabular}
\end{table}

\begin{figure}[htb]
\centering
\includegraphics[width = 0.9\linewidth]{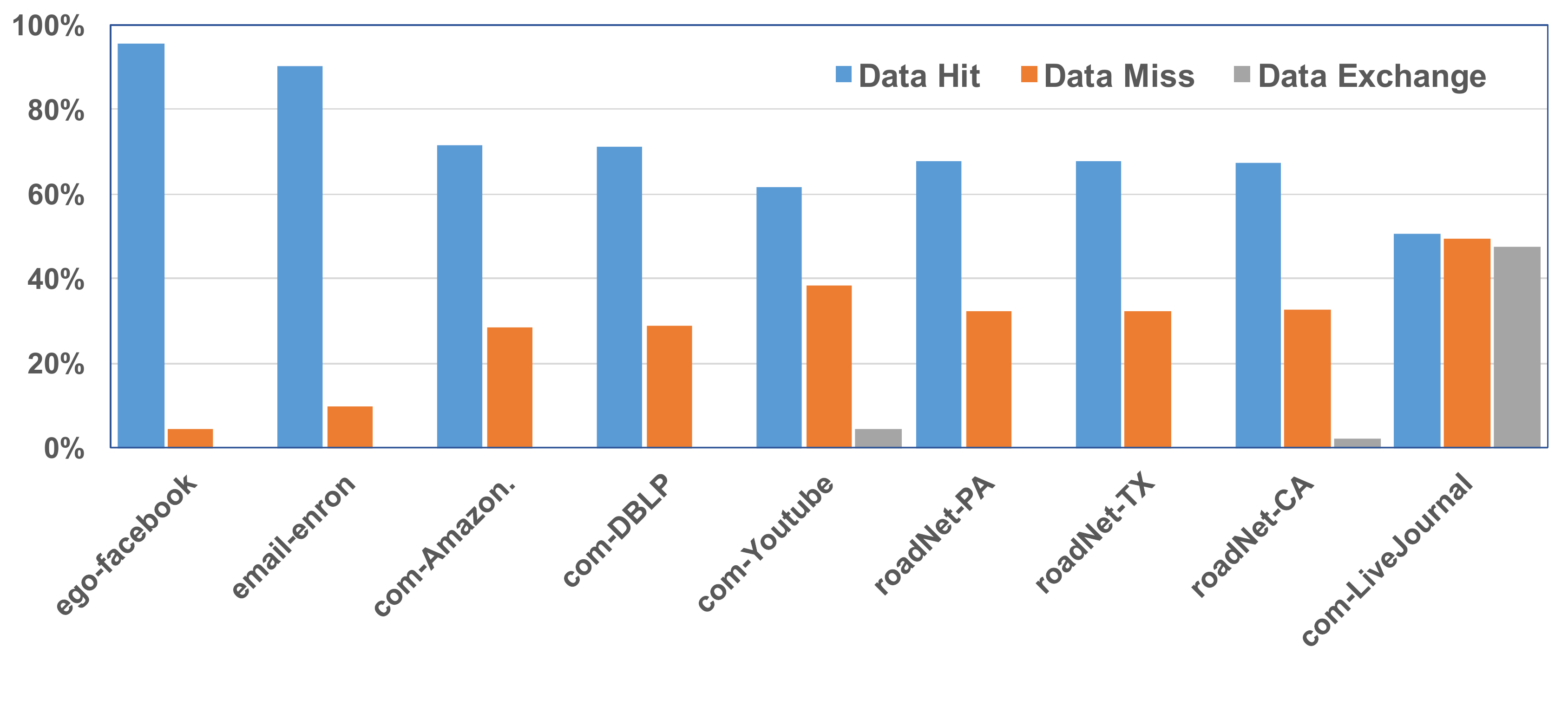}
\caption{Percentages of data hit/miss/exchange.}
\label{fig:datareuse}
\end{figure}

When the STT-MRAM computational memory size is smaller than those listed in TABLE~\ref{tab:sliceDataSize}, data exchange will happen. For example, with $16$ MB, the three large graphs will have to do data exchange as shown in Fig.~\ref{fig:datareuse}. In this figure, we also list the percentages of data hit (average $72\%$) and data miss (average $28\%$). Recall that the first time a data slice is loaded, it is always a miss, and a data hit implies that the slice data has already been loaded. So this shows that the proposed data reuse strategy saves on average $72\%$ memory \texttt{WRITE} operations.

\subsection{Benefits of Data Slicing}

As shown in TABLE~\ref{tab:validSlice}, the average percentage of valid slices in the five largest graphs is only $0.01\%$. Therefore, the proposed data slicing strategy could significantly reduce the needed computation by $99.99\%$.

\begin{table}[htbp]
\setlength{\tabcolsep}{4pt}
\caption{Percentage of valid slices.}
\label{tab:validSlice}
\centering
\begin{tabular}{lr||lr||lr}
\specialrule{0.8pt}{0pt}{0pt}
ego-facebook & 7.017\% & com-DBLP & 0.036\% & roadNet-TX & 0.010\% \\
email-enron & 1.607\% & com-Youtube & 0.013\% & roadNet-CA & 0.007\% \\
com-Amazon & 0.014\% & roadNet-PA & 0.013\% & com-lj & 0.006\% \\
\specialrule{0.8pt}{0pt}{0pt}
\end{tabular}
\end{table}

\subsection{Performance and Energy Results}

TABLE~\ref{tab:graphperf} compares the performance of our proposed in-memory TC accelerator against a CPU baseline implementation, and the existing GPU and FPGA accelerators.
One can see a dramatic reduction of the execution time in the last columns from the previous three columns. Indeed, without PIM, we achieved an average $53.7\times$ speedup against the baseline CPU implementation because of data slicing, reuse, and exchange. With PIM, another $25.5\times$ acceleration is obtained.
Compared with the GPU and FPGA accelerators, the improvement is $9\times$ and $23.4\times$, respectively. It is important to mention that we achieve this with a single-core CPU and $16$ MB STT-MRAM computational array.

\begin{table}[htbp]
\setlength{\tabcolsep}{5pt}
\caption{Runtime (in seconds) comparison among our proposed methods, CPU, GPU and FPGA implementations.}
\label{tab:graphperf}
\centering
\begin{tabular}{l|r|r|r|r|r}
\specialrule{0.8pt}{0pt}{0pt}
 \multirow{2}{*}{Dataset} & \multirow{2}{*}{CPU} & \multirow{2}{*}{GPU \cite{XiongTCFPGA}} & \multirow{2}{*}{FPGA \cite{XiongTCFPGA}} & \multicolumn{2}{c}{This Work}\\ \cline{5-6}
 & & & & w/o PIM & TCIM \\ \hline
 ego-facebook & 5.399 & 0.15 & 0.093 & 0.169 & 0.005 \\
 email-enron & 9.545 & 0.146 & 0.22 & 0.8 & 0.021 \\
 com-Amazon & 20.344 & N/A & N/A & 0.295 & 0.011 \\
 com-DBLP & 20.803 & N/A & N/A & 0.413 & 0.027 \\
 com-Youtube & 61.309 & N/A & N/A & 2.442 & 0.098 \\
 roadNet-PA & 77.320 & 0.169 & 1.291 & 0.704 & 0.043 \\
 roadNet-TX & 94.379 & 0.173 & 1.586 & 0.789 & 0.053 \\
 roadNet-CA & 146.858 & 0.18 & 2.342 & 3.561 & 0.081 \\
 com-LiveJournal & 820.616 & N/A & N/A & 33.034 & 2.006 \\
\specialrule{0.8pt}{0pt}{0pt}
\end{tabular}
\end{table}

\begin{figure}[htbp]
\centering
\includegraphics[width = 1.0\linewidth]{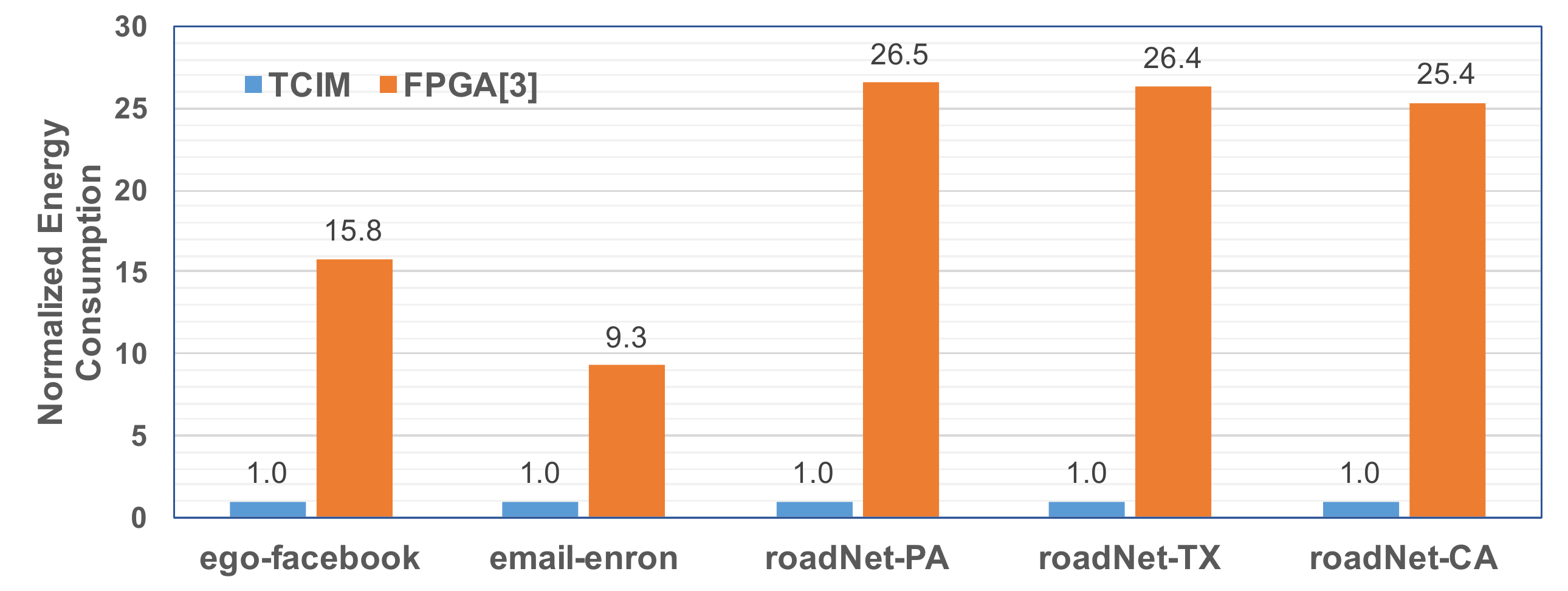}
\caption{Normalized results of energy consumption for TCIM with respect to FPGA.}
\label{fig:energy}
\end{figure}

As for the energy savings, as shown in Fig.~\ref{fig:energy}, our approach has $20.6\times$ less energy consumption compared to the energy-efficient FPGA implementation \cite{XiongTCFPGA}, which benefits from the non-volatile property of STT-MRAM and the in-situ computation capability.

\section{Conclusion}\label{sec:conclusion}

In this paper, we propose a new triangle counting (TC) method, which uses massive bitwise logic computation, making it suitable for in-memory implementations.
We further propose a sparsity-aware processing-in-MRAM architecture for efficient in-memory TC accelerations: by data slicing, the computation could be reduced by $99.99\%$, meanwhile the compressed graph data can be directly mapped onto STT-MRAM computational memory array for bitwise operations, and the proposed data reuse and exchange strategy reduces $72\%$ of the memory \texttt{WRITE} operations.
We use device-to-architecture co-simulation to demonstrate that the proposed TC in-memory accelerator outperforms the state-of-the-art GPU and FPGA accelerations by $9\times$ and $23.4\times$, respectively, and achieves a $20.6\times$ energy efficiency improvement over the FPGA accelerator.

Besides, the proposed graph data compression and data mapping strategies are not restricted to STT-MRAM or TC problem. They can also be applied to other in-memory accelerators with other non-volatile memories.

\bibliographystyle{unsrt}

\scriptsize

\begingroup

\endgroup

\end{document}